\documentclass[a4paper,11pt]{article}
\pdfoutput=1 

\usepackage{jheppub} 

\usepackage[T1]{fontenc} 
\usepackage{caption}
\usepackage{comment}
\usepackage{amsmath}
\usepackage{graphicx}
\usepackage{tikz}
\usetikzlibrary{arrows,decorations.pathmorphing}
\usepackage[compat=1.1.0]{tikz-feynman}
\usepackage{amssymb}
\usepackage{bm,commath,mathtools,wasysym,xfrac}
\usepackage{pdfpages}
\usepackage{tcolorbox}
\usepackage{braket}
\usepackage{MnSymbol}
\usepackage{cancel}
\usepackage[english]{babel} 
\usepackage{blindtext}

\usepackage{cancel}
\usepackage{color}

\title{\boldmath Large deflection scattering, soft radiation and KMOC formalism }

\author[a,b,c]{Samim Akhtar,}
\author[d]{Alok Laddha,}
\author[a,b,e]{Arkajyoti Manna,}
\author[a,b]{and Akavoor Manu}

\affiliation[a]{The Institute of Mathematical Sciences \\
 	IV Cross Road, C.I.T. Campus, Taramani, Chennai 600 113, India}
 \affiliation[b]{Homi Bhabha National Institute \\
	Training School Complex, Anushakti Nagar, Mumbai 400 094, India}
   \affiliation[c]{ICTP South American Institute for Fundamental Research, Instituto de Física Teórica UNESP \\
  Rua Dr. Bento Teobaldo Ferraz 271, 01140-070, São Paulo, SP, Brazil} 
 \affiliation[d]{Chennai Mathematical Institute \\ H1, SIPCOT IT Park, Siruseri, Kelambakkam 603103, India}
 \affiliation[e]{Center for High Energy Physics, Indian Institute of Science,\\ C.V. Raman Avenue,
	Bangalore 560012, India}
\emailAdd{samim.akhtar@ictp-saifr.org}
\emailAdd{aladdha@cmi.ac.in}
\emailAdd{arkajyotim@iisc.ac.in}
\emailAdd{amanu@imsc.res.in}

\abstract{KMOC (Kosower, Maybee, and O'Connell) formalism is an approach to analyze classical scattering in gauge theories and gravity using a class of ``inclusive'' observables which can be computed solely from on-shell amplitudes \cite{Kosower:2018adc}. This formalism has led to striking developments in the context of perturbative scattering, which corresponds to large impact parameter scattering. As a result, in its current form, the KMOC formulae cannot be directly applied to processes for generic values of the impact parameter. 

However, there is a domain where the relationship between classical radiation and on-shell amplitudes can be stretched beyond large impact parameter scattering. This regime is defined by the soft expansion of outgoing radiation. It is thus natural to ask whether such soft radiative fields can be computed using the basic paradigm set by the KMOC formalism. In this short note, we show that this is indeed the case for electromagnetic memory. In particular, we compute an inclusive observable associated with soft flux at ${\cal I}^{+}$
 and show that, irrespective of the details of the hard scattering, this observable defines a non-perturbative formula for the electromagnetic memory in the classical limit. We argue that the result obtained for electromagnetic memory using the KMOC paradigm is consistent with that of \cite{Laddha:2018rle}, where the classical limit of the quantum soft theorem was derived using saddle-point analysis. The gravitational case, however, is qualitatively different due to the presence of the nonlinear memory effect, which requires knowledge of the hard amplitude. Consequently, unlike the electromagnetic memory, we have not been able to show consistency of the leading soft graviton theorem and the soft inclusive gravitational flux obtained using the KMOC formalism.}

\makeatletter
\gdef\@fpheader{}
\makeatother

\begin{document} 
	\maketitle
	\flushbottom

\section{Introduction}
In recent years, there has been a surge of activity in the analysis of classical scattering in gauge theories and gravity \cite{Bern:2021yeh,Bern:2021xze,Bern:2023ccb,Kalin:2019rwq,Kalin:2019inp,Cho:2021arx,Dlapa:2021vgp,Dlapa:2022lmu,DiVecchia:2021bdo,Bjerrum-Bohr:2021din,Dlapa:2025biy,Bini:2022enm,Bini:2020rzn,Bini:2024rsy,Bern:2019crd,Bern:2020buy,Akhtar:2024mbg,Bini:2024pdp,Arkani-Hamed:2019ymq,Bautista:2024emt,Aoki:2024boe,Chen:2022clh,Akhtar:2024lkk,Foffa:2016rgu,Vitale:2014mka,Vines:2017hyw,Luna:2023uwd,Manu:2020zxl,Georgoudis:2025vkk} using on-shell amplitudes. Based on a series of earlier works \cite{Neill:2013wsa,Bjerrum-Bohr:2013bxa,Cachazo:2017jef,Donoghue:1994dn,Donoghue:1996mt,Holstein:2004dn,Damour:2016gwp,Cheung:2018wkq}, a first-principle approach for computing the final state of a classical scattering process from quantum S-matrix was proposed by Kosower, Maybee, and O'Connell (KMOC) in a seminal work \cite{Kosower:2018adc}. KMOC proved that the final state of a classical electromagnetic (or gravitational) scattering is obtained from a class of inclusive observables that depend on the square of the S-matrix and the specification of an initial state. These observables compute the conditional expectation of an observable in the far future, given a semi-classical state in the far past. More in detail, when the initial state is a one-parameter family of two-particle coherent states\footnote{We use the term ``coherent state'' to refer to a state with two particles whose momenta are peaked around classical values and are described by gaussian wave packets.}(with the parameter specifying the impact parameter of the scattering), the resulting inclusive observable peaks around a classical value when we take the exchange momenta to zero in the classical limit. Hence, a two-body hyperbolic scattering in gravity or electromagnetism can be analyzed in terms of the classical limit of a class of quantum observables.  In fact, in \cite{Caron-Huot:2023vxl}, it was realized that the inclusive observables that are computed using the KMOC formalism are a special class of OTOC (out of time ordered) observables that can be computed in terms of monomials of the S-matrix. These observables were referred to as $\textrm{Exp}_{3}$ observables \cite{Caron-Huot:2023vxl}\footnote{We use the term $\textrm{Exp}_{3}$ in a broader sense than reference~[40], where it refers specifically to the one-point asymptotic field observable. Here we use it to denote the general class of inclusive observables of the form $\langle S^\dagger \mathcal{O} S\rangle$ computed in the KMOC formalism, including the impulse, radiated momentum, and radiative flux.}.

The assumption of a large impact parameter is rather crucial for the KMOC formalism, as the classical limit is parametrized in terms of the soft limit of exchange momenta. The softness of exchange momenta (or equivalently, large impact parameter scattering) implies that given an incoming coherent state of matter particles, the classical observables in the far future, such as final momenta of the scattering particles, can be obtained perturbatively. Hence, in the case of electromagnetic and gravitational scattering, the KMOC formalism is particularly well suited for computing quantities such as the radiation flux in the so-called post-Lorentzian (PL) expansion \cite{Bern:2021xze, Saketh:2021sri} and the post-Minkowskian (PM) expansion, respectively\footnote{The PL expansion is an expansion in the electromagnetic coupling and is directly analogous to the post-Minkowskian (PM) expansion in gravity.}.

The generic classical scattering problem is not integrable, and hence we expect that computation of asymptotic observables such as linear impulse or radiated momentum at future null infinity must rely on perturbation theory. This is where soft radiation measured by a detector in the far (retarded) future offers an interesting counterpoint to the usual perturbative analysis. This is mainly due to the following reasons 
\begin{itemize}
\item Electromagnetic (gravitational) memory, also known as the leading soft mode of the radiative field,  and a set of logarithmic tail terms in the soft expansion, are exact to all orders in the PL (PM) expansion and are universal in the sense that they are independent of the details of the irrelevant operators present in the Lagrangian. 
\item The perturbative exactness and universality of soft modes, such as the memory, are intimately tied to the fact that the S-matrix in QED (gravity) satisfies a hierarchy of non-perturbative soft photon (graviton) theorems. 
\item As the only input in the $\textrm{Exp}_{3}$ observables, which are computed in the KMOC formalism, is the incoming state, the final momenta of scattering particles are computed perturbatively, and this leads to a perturbative computation of soft radiation including the memory effect. Thus, the power of the quantum soft theorem, its universality, and non-perturbative aspect are underutilized in the KMOC formalism. This is reflected in the fact that while non-perturbative proof of classical soft theorems exist \cite{Saha:2019tub,Ghosh:2021bam,Sahoo:2020ryf,Sahoo:2021ctw,Karan:2025ndk} and have led to impressive advances including analytic results for all order soft theorem in electromagnetic scattering, memory as an $\textrm{Exp}_{3}$ observable can only be computed perturbatively\footnote{In \cite{Bautista:2021llr}, it was shown that the non-perturbative soft theorems can be used to constrain a large class of ``moments'' which are computed in KMOC formalism.}.
\end{itemize}
However, in \cite{Laddha:2018rle}, A. Sen and one of the present authors proved that one could analyze the classical limit of quantum soft factorization theorems as follows: Given a set of incoming and outgoing particles whose charges and asymptotic momenta are fixed, one can compute the probability of $N$ photons (or gravitons) being emitted in a specific bin whose range of frequencies is centered around $\omega$ which is soft compared to characteristic frequency of the system. By extremizing the number of photons (gravitons) that reach the given bin, classical soft radiation emitted in that bin was computed. This approach relied on the existence of multi-soft factorization theorems and showed that a class of observables could be computed directly using in-out asymptotics and soft theorems.

In this paper, we try to compare and contrast these two approaches aimed at computing classical soft radiation using quantum soft theorems and argue that the $\textrm{Exp}_{3}$ observables can be used to compute electromagnetic memory in four dimensions regardless of our knowledge of the hard scattering. In a nutshell, we consider a class of two-particle incoming state, which is a coherent state in the limit of vanishing impact parameter, and show that, irrespective of the details of the hard amplitude, soft factorization is sufficient to compute classical electromagnetic memory via the in-in formalism. We show how the number of photons that contribute to the classical limit is not fixed by the order of perturbative expansion but is, in fact, extremized just as in \cite{Laddha:2018rle}. For gravity, however, the situation is more subtle due to the presence of the nonlinear memory effect, which necessitates knowledge of the hard amplitude. 

The paper is organized as follows. In section \ref{sec:kmoc}, we review the computation of electromagnetic radiation using the KMOC formalism. We take this opportunity to bring out certain salient aspects of the formalism that, in our opinion, are rather implicit in the literature. In section \ref{sec:classical_extremal}, we review the derivation of classical memory effect to all orders in perturbative expansion using multi-soft theorems \cite{Laddha:2018rle}. In section \ref{main_sec}, we then argue that starting with an initial state which violates the second inequality of the Goldilocks hierarchy, one can still compute the electromagnetic memory, and this result is consistent with the result obtained in \cite{Laddha:2018rle}. In section \ref{gravity_sec}, we extend this derivation to compute the gravitational memory. We finish with some concluding remarks.

\section{Review of KMOC formalism} \label{sec:kmoc}
In this section, we review the KMOC formalism. This is an approach to compute classical observables, such as the radiative flux of electromagnetic (or gravitational) field, using scattering amplitudes.  Our review focuses on the computation of radiative flux using this formalism, and at the risk of writing a longer than necessary review, we elaborate on certain aspects of the derivation that bring out some subtle aspects involved in taking the classical limit.  

The classical processes that can be analyzed in this approach so far include two classical particles undergoing large impact parameter scattering. This leads to the formulation where sources can be modelled via effective field theory techniques and their scattering analyzed systematically using on-shell scattering amplitudes methods. 

Naturally, the perturbation parameter in this case is the inverse of the impact parameter. So, the essential idea of the formalism is to use coherent states to model the initial classical matter particles (with potentially infinitely many multipole moments)  and compute the expectation value of an operator such as the graviton energy-momentum operator at future null infinity ${\cal I}^{+}$ with the Heisenberg evolution of the observable governed by the S-matrix. The classical limit is then taken by realizing that in the case of small-deflection scattering, the exchange momenta, as well as momenta of radiated photons, are vanishingly small.

We shall concern ourselves with the scattering of two massive classical scalar particles. So the initial state is the coherent state generated by a two-particle Fock state
\begin{align} \label{in_state}
    \ket{\text{in}}_{b} &= \int \prod_{i=1}^{2}\ d\Phi(p_{i}) \phi_{i}(p_{i})\ e^{\frac{ib\cdot p_1}{\hbar}}\ \ket{p_{1},p_{2}}\,,\cr
    \braket{\hat{P}^{\mu}_{i}} &= m_{i}u^{\mu}_{i} + \mathcal{O}(\hbar)\,,\cr 
    \braket{\hat{P}_{i}^{\mu}\hat{P}_{i}^{\nu}} & - \braket{\hat{P}_{i}^{\mu}}\braket{\hat{P}_{i}^{\nu}} = \mathcal{O}(\hbar) \,\ \ \forall i=1,2,
\end{align}
where $\phi_{i}(p_{i})$s are the Gaussian wavepackets of the two particles, $m_{i}u^{\mu}_{i}$ are the classical momenta around which the Gaussians are peaked and $b^{\mu}$ is the impact parameter, which sets the characteristic length scale in the problem. The measure $d\Phi(p_i)$ is defined as follows
\begin{align} 
    d\Phi(p_i):=\hat{d}^4p_i \hat{\delta}^{(+)}(p_i^2-m_i^2)\, ,\  \hat{d}^4p_i:=\frac{d^4p_i}{(2\pi)^4}\,,\  \hat{\delta}^{(+)}(p_i^2-m_i^2):=2\pi\, \Theta(p_i^0)\delta(p_i^2-m_i^2) \,.
\end{align}
Phenomenologically, one of the most important observables is the radiation flux emitted during a classical scattering process, and so, we shall concentrate on this observable. The operator whose classical limit gives the radiative flux is the energy-momentum operator for that particular massless quanta.

Since the in-state is composed solely of classical matter particles, the radiated energy and momentum are defined as
\begin{equation}\label{rad_form}
    {\cal K}^{\mu} = \bra{\text{out}}\hat{K}^{\mu}\ket{\text{out}}=\bra{\text{in}}_{b}S^{\dagger}\hat{K}^{\mu}S\ket{\text{in}}_{b} = \bra{\text{in}}_{b}T^{\dagger}\hat{K}^{\mu}T\ket{\text{in}}_{b},
\end{equation}
where $S=I+iT$ and 
\begin{align}
\hat{K}^{\mu}\, =\, \sum_{h}\, \int d\Phi(k)\ k^{\mu} a_{h}^{\dagger}(k)\, a_{h}(k)\ ,
\end{align}
where $h$ is the helicity of the quanta. The expression for the radiative flux carried by $X$ quanta with momenta $k_j$ ($1\leq j \leq N$) can be written as
\begin{equation}\label{rad_photon_N}
     {\cal K}^{\mu} = \sum_{X=1}^{\infty}\sum_{R}\int d\Phi (R) d\Phi(X) \bra{\text{in}}_{b}T^{\dagger}\ket{R,X}(k_{1}+\ldots +k_{X})^{\mu}\bra{R,X}T\ket{\text{in}}_{b}\ ,
\end{equation}
where the $\sum_{X}$ denotes the contribution from an arbitrary number of massless particles. Here $d\Phi(X)$ is the phase space measure over the set of all the momenta of the massless particles in $X$, defined by
\begin{align}
    d\Phi (X) :=\prod_{j=1}^X d\Phi_j(k)= \prod_{j=1}^X \hat{d}^4 k_j \hat{\delta}^{(+)}(k_j^2) \,.
\end{align}
For pedagogy, we have suppressed the sum over the helicities of the massless particles. The $\sum_{R}$ denotes the contribution from the other quanta present in the theory. The observable defined in eq.\eqref{rad_photon_N} can be used to compute the radiation flux for any scattering process. To understand how the above is used to compute the classical radiative flux for large impact parameter scattering, we shall restrict ourselves to a particular setup. Consider the classical small deflection scattering of two charged massive scalar particles modeled via scalar QED. In this case, the initial state defined in eq.\eqref{in_state} will also have charges that is
\begin{equation}
    \ket{\text{in}}_{b} = \int \prod_{i=1}^{2}\ d\Phi(p_{i}) \phi_{i}(p_{i})\ e^{\frac{ib\cdot p_1}{\hbar}}\ \ket{p_{1},Q_{1};p_{2},Q_{2}}.
\end{equation}
The radiative flux is computed for electromagnetic waves by taking the expectation value of the photon energy–momentum operator. The radiative flux after some manipulation becomes
\begin{equation}\label{rad_LIS}
    \begin{split}
        {\cal K}^{\mu} & = \sum_{X}\int d\Phi(r_{1}) d\Phi(r_{2}) d\Phi(X) \bra{\text{in}}_{b}T^{\dagger}\ket{r_{1},r_{2},X}(k_{1}+\ldots +k_{X})^{\mu}\bra{r_{1},r_{2},X}T\ket{\text{in}}_{b}\\
        &=\sum_{X=1}^{\infty}\frac{1}{X!}\int\prod_{i=1}^{X}d\Phi(k_{i})(\sum_{j=1}^{X}k_{j}^{\mu})\\
                             &\bigg| \int\prod_{i=1}^{2}\hat{d}^{4}q_{i}
                             \delta(2p_{i}\cdot q_{i}+q_{i}^{2})\ e^{-iq_{1}\cdot b/\hbar}\\
                             &\times\delta^{(4)}(q_{1}+q_{2}-k_{1}-\ldots-k_{N})A_{4+X}(p_{1},p_{2},k_{1}\ldots,k_{X}\rightarrow p_{1}+q_{1},p_{2}+q_{2})\bigg|^{2}\ .
    \end{split}
\end{equation}

We can now discuss the classical scalings involved. Examining the above expression, we see that we can interpret the $q^{\mu}_{i}$s as the exchange momenta. In the $\hbar\rightarrow 0$, however, the exponential becomes highly oscillatory unless $|\vec{q}_{1}|\rightarrow 0$ while $b^{\mu}$ remains fixed\footnote{We could have started with an initial state with $b^{\mu}_{1}$ and $b_{2}^{\mu}$, then the same argument would hold for $q_{2}$ as well. In this case the impact parameter would be $b^{\mu} = b_{1}^{\mu}-b_{2}^{\mu}$. For simplicity we have taken $b^{\mu}_{2}=0$.}. So, we see that the $q\sim 0$ region is the one which contributes to classical scattering\footnote{We could also have taken $|\vec{b}|\rightarrow 0$ while keeping $q$ fixed. We shall discuss this possibility at the end of this section.}. To compute the contribution from this region of $q$-space in $\hbar\neq 1$ units, we go over to wave number space i.e $q_{i}^{\mu}=\hbar\bar{q}_{i}^{\mu}$. The momentum-conserving delta functions imply that the outgoing momenta of the photons will also scale in the same manner, i.e, $k_{i}^{\mu}=\hbar\bar{k}_{i}^{\mu}$. However, these are not all the massless quanta that we have to account for. In perturbation theory, loops will contribute to the classical scattering, but since the loop momenta are not constrained by the external momenta, their scaling doesn't follow from the above arguments.\\
To see how the loop momenta scale, we require the scaling of the stripped amplitude, $A_{4+X}(p_{1},p_{2},k_{1}\ldots,k_{X}\rightarrow p_{1}+q_{1},p_{2}+q_{2})$. For finite frequency radiation, this is complicated to obtain, so we shall use soft factorization to obtain the required scaling. 
Let us restrict the range of the frequency of the photons between $[\omega,\omega+\delta \omega$], with $\omega \ll |\vec{p}_{i}|$, where $\vec{p}_{i}$'s denote the classical momenta of the initial particles. Since the classical momenta of the outgoing particles are perturbatively constructed from their initial values, $\omega$ will also be small compared to the final momenta of the particles. In the soft limit, the five-point amplitude factorizes as follows \cite{Low:1958sn,Weinberg:1965nx,Laddha:2018myi}
\begin{equation}
    \mathcal{A}_{5}(p_{1},p_{2},k\rightarrow p'_{1},p'_{2})|_{\omega\rightarrow 0} = \frac{1}{\omega}S^{(0)}\times \mathcal{A}_{4}(p_{1},p_{2}\rightarrow p'_{1},p'_{2})  \ ,
\end{equation}
where the amplitudes are the unstripped amplitudes and
\begin{equation}
    S^{(0)} = -\sum_{i=1}^{2} Q_{i}\frac{\varepsilon(k)\cdot p_{i}^\prime}{p_{i}^\prime \cdot k} + \sum_{j=1}^{2} Q_{j}\frac{\varepsilon(k)\cdot p_{j}}{p_{j} \cdot k}
\end{equation}
and $k^{\mu} = \omega(1,\hat{n})$ where $\hat{n}$ is the direction of the soft photon. Since the above is a multiplicative factor, the factorisation for the $A_{4+X}$ amplitude is just the $A_{4}$ amplitude multiplied with $X$ factors of $S^{(0)}$\footnote{We note that for illustrating the classical scaling for loop momenta we are working in scalar QED but the soft factorisation in the above equation is universal since the soft factor only depends on the momenta of the external particles.}. In this case, the momentum-conserving delta function in the last line of eq.\eqref{rad_LIS} is independent of the soft momenta, so we get
\begin{equation}\label{soft_rad_flux1}
    \begin{split}
        {\cal K}_{\text{soft}}^{\mu} &=\sum_{X=1}^{\infty}\frac{1}{X!}\int_{\omega}^{\omega+\delta\omega}\prod_{i=1}^{X}d\Phi(k_{i})(\sum_{j=1}^{X}k_{j}^{\mu})\\
                             &\bigg| \int\prod_{i=1}^{2}\hat{d}^{4}q_{i}
                             \delta(2p_{i}\cdot q_{i}+q_{i}^{2})\ e^{-iq_{1}\cdot b/\hbar}\\
                             &\times\delta^{(4)}(q_{1}+q_{2})\times (S^{(0)})^{X}\times\ A_{4}(p_{1},p_{2}\rightarrow p_{1}+q_{1},p_{2}+q_{2})\bigg|^{2}\ .
    \end{split}
\end{equation}
 We use this formula to show that in the classical limit, the region of integration in the space of loop momenta that survives is a ball centered around the origin, whose radius scales with $\hbar$. Let us now look at the one photon emission contribution to eq.\eqref{soft_rad_flux1}
\begin{align}
    {\cal K}^{(1)\mu}_{\text{soft}} = \int_{\omega}^{\omega+\delta\omega} d\Phi(k)\ k^{\mu}\ \bigg| \int\prod_{i=1}^{2}\hat{d}^{4}q_{i}
                             \delta(2p_{i}\cdot q_{i}+q_{i}^{2})\ & e^{-iq_{1}\cdot b/\hbar}
                              \delta^{(4)}(q_{1}+q_{2}) \cr
                             & \hspace{{-1cm}}S^{(0)}\ A_{4}(p_{1},p_{2} \rightarrow p_{1}+q_{1},p_{2}+q_{2})\bigg|^{2} \,.
\end{align}
where the superscript denotes the contribution from the one-photon emission. The coupling expansion now is contained in $A_{4}$, and at $1$-loop, there are multiple topologies to consider. One of them is the box topology, and the expression for the same is
\begin{equation}
    iB = (Q_{1}Q_{2})^{2}\int \hat{d}^{4}l \frac{(2p_{1}+l)\cdot (2p_{2}-l)(2p_{1}+l+q)\cdot (2p_{2}-l-q)}{(2p_{1}\cdot l +l^{2}+i\epsilon)(-2p_{2}\cdot l +l^{2}+i\epsilon)} \ .
\end{equation}
 We now know that for large impact parameter scattering, the $q\sim 0$ region contributes. So, we can divide the integration region into $l>>q$ and $l\sim q$. To compute the contribution from the latter region, we again go to wavenumber space $l^{\mu}=\hbar\bar{l}^{\mu}$. For the former region, we note from the above expression that
\begin{equation}
    iB = (Q_{1}Q_{2})^{2}\int_{l>>q}\ \hat{d}^{4}l \frac{(2p_{1}+l)\cdot (2p_{2}-l)(2p_{1}+l)\cdot (2p_{2}-l)}{(2p_{1}\cdot l +l^{2}+i\epsilon)(-2p_{2}\cdot l +l^{2}+i\epsilon)}
+ \mathcal{O}(q) \ .
\end{equation}
The above expression doesn't have any $q$ dependence. This means that when we plug this expression into the previous expression and take the classical limit, we get
\begin{align}
   {\cal K}^{(1)\mu}_{\text{soft}}= \hbar^{2}(Q_{1}Q_{2})^{2}\int_{\omega}^{\omega+\delta\omega} & d\Phi(\bar{k}) \  \bar{k}^{\mu}\  \bigg| \int\hat{d}^{4}\bar{q}
                             \prod_{i=1}^{2}\delta(2p_{i}\cdot \bar{q})\ e^{-i\bar{q}\cdot b} \cr
                             &\times
                             \ \bigg(\sum_{i=1}^{2}(-1)^{i}\ Q_{i}\bigg[\frac{\varepsilon\cdot \bar{q}}{p_{i}\cdot \bar{k}} -(\varepsilon\cdot p_{i})\frac{\bar{q}\cdot \bar{k}}{(p_{i}\cdot \bar{k})^{2}}\bigg]\bigg) \ B(p_{1},p_{2})\bigg|^{2} \,.
\end{align}
Now the $q$ integral can be easily done and we obtain terms which are polynomial in derivatives (w.r.t $b^{\mu}$) of $\delta^{(2)}(\vec{b})$. These terms don't contribute to large impact parameter scattering as $b$ is finite. We conclude that the $l\gg q$ region doesn't contribute to large impact parameter scattering and only the $l\sim q$ region contributes. This conclusion holds for all topologies and for all loop momenta. So, for loop momenta $l_{i}^{\mu}=\hbar\bar{l}^{\mu}_{i}$. So, we see that the scaling is determined by the region in $q,l$ and $k$-space which dominates the scattering process, and in this case, \textit{all the massless momenta scaled with $\hbar$ in the classical limit}. Finally, in $\hbar\neq 1$ units, the charges are not dimensionless and $e\rightarrow e/\sqrt{\hbar}$ and they also contribute additional factors of $\hbar$\footnote{This follows from the fact that the fine structure constant $\alpha = \frac{e^{2}}{\hbar}$ is dimensionless.}.  Equipped with the amplitudes and doing the scaling, it is a straightforward exercise to compute the classical radiation. The radiation at leading order in the small angle has been computed, for electromagnetism, using this procedure in \cite{Kosower:2018adc}. Even though we have described the classical scalings for massive charged scalar particles, since the scalings are determined by the dominant regions of $l,q$, and $k$-space, these will also hold for particles with higher multipole moments. We emphasize that the use of the soft limit here is purely a simplifying device; the conclusion that loop momenta scale as $l^\mu \sim \hbar$ in the classical limit follows from the scaling of external momenta alone and extends to finite-frequency radiation.

We now highlight a few aspects of the formalism that will be relevant for the later sections.
\begin{enumerate}
\item  The scaling of massless momenta ensures that the KMOC formalism is best suited for computing observables such as radiation at future null infinity in large impact parameter scattering. The specification of a semi-classical in-state then (under evolution) leads to a (semi-classical) out-state where the momenta are peaked around $p_{i} + O(\frac{1}{\vert b\vert})$ as $\vert b\vert\, \rightarrow\, \infty$. Thus, the computation of classical observables via the KMOC formalism integrates beautifully with perturbative schemes such as Post-Minkowskian methods used to analyse gravitational scattering.  
\item Since, for small-angle scattering (large impact parameter), the out-state is perturbatively obtained from the initial state, the energy radiated away in the radiation flux will also be small compared to the energies of the initial matter particles. Nonetheless, it may appear counterintuitive that the dominant contribution to the radiative flux arises from a single photon emission in the quantum amplitude. This is equivalent to the statement that $\langle\textrm{out}\, \vert\, \hat{F}_{\mu\nu}(\omega, \hat{n})\, \vert\, \textrm{out} \rangle$ does not vanish when the in-state is a two-particle matter coherent state. In fact, it turns out that the dominance of single photon emission in the classical limit is once again associated with the fact that the KMOC formalism deals with perturbative (large impact parameter) scattering. 
To verify this statement, we shall again compute the soft radiation flux using eq.\eqref{soft_rad_flux1} because in this setting the scalings of the amplitudes are simpler to obtain. So, we start with the earlier formula in eq.\eqref{soft_rad_flux1}
\begin{equation}\label{radiation}
\begin{split}
&\mathcal{K}^{\mu}_{\text{soft}} = \int\prod_{i=1}^{2}\hat{d}^{4}q_{i}
                             \delta(2p_{i}\cdot q_{i}+q_{i}^{2})\hat{d}^{4}w_{i}
                             \delta(2p_{i}\cdot w_{i}+w_{i}^{2})\ e^{-iq_{1}\cdot b/\hbar}\delta^{(4)}(q_{1}+q_{2})\delta^{(4)}(w_{1}+w_{2})\\
                             &A^{*}_{4}(p_{1}+w_{1},p_{2}+w_{2}\rightarrow p_{1}+q_{1},p_{2}+q_{2})A_{4}(p_{1},p_{2}\rightarrow p_{1}+w_{1},p_{2}+w_{2})\\
    &\bigg[ \int_{\omega}^{\omega+\delta\omega} d\Phi(k)\ k^{\mu}\\
    &\ S_{k}^{*(0)}(p_{1}+w_{1},p_{2}+w_{2}\rightarrow p_{1}+q_{1},p_{2}+q_{2})S_{k}^{(0)}(p_{1},p_{2}\rightarrow p_{1}+w_{1},p_{2}+w_{2}) +\ldots\bigg] ,
\end{split}
\end{equation}
where the sum over helicity of the on-shell photons has been suppressed. The `$\ldots$' contain contributions from more than one photon emission.
For large impact parameter scattering in electrodynamics, the first non-trivial order at which radiation is emitted is $e^{6}$ as the $4$-point amplitude in the second line of the above expression, would be just a tree-level amplitude. To see this, we take the classical limit of the above expression and the following classical expansions\footnote{It should be understood that whenever factors of $\hbar$ appear, the exchange momenta are to be interpreted as wavenumbers.} 
\begin{equation}
    A^{\text{tree}}_{4}(1,2\rightarrow 1+q,2-q)|_{\hbar\rightarrow 0} = \frac{Q_{1}Q_{2}e^{2}}{\hbar^{3}}\frac{(4p_{1}\cdot p_{2})}{q^{2}} +\mathcal{O}(\hbar^{-1})
\end{equation}
and the leading soft factor has the following expansion
\begin{equation}
\begin{split}
    S^{(0)}|_{\hbar\rightarrow 0} &= \frac{Q}{\sqrt{\hbar}}\frac{1}{\hbar}\bigg( \frac{(p+\hbar q)^{\mu}}{p\cdot k} - \frac{p^{\mu}}{p\cdot k}\bigg)\\
    &= \frac{Q}{\sqrt{\hbar}}\bigg[\frac{1}{\hbar}\frac{p^{\mu}}{p\cdot k}\bigg(\sum_{n=1}^{\infty}(-1)^{n}\hbar^{n}\bigg(\frac{q\cdot k}{p\cdot k}\bigg)^{n}\bigg) + \frac{q^{\mu}}{p\cdot k}\bigg(1+\sum_{n=1}^{\infty}(-1)^{n}\hbar^{n}\bigg(\frac{q\cdot k}{p\cdot k}\bigg)^{n}\bigg)\bigg]\\
    &= \frac{Q}{\sqrt{\hbar}}\ (\bar{S}^{(0)} +\hbar \bar{S}_{1}^{(0)} +\hbar^{2}\bar{S}_{2}^{(0)} +\ldots)\\
    & = \frac{Q}{\sqrt{\hbar}}\ \bigg(-\frac{p^{\mu}}{p\cdot k}\bigg(\frac{q\cdot k}{p\cdot k}\bigg)+\frac{q^{\mu}}{p\cdot k}\bigg) +\mathcal{O}(\sqrt{\hbar})\ .
\end{split}
\end{equation}
With these classical expansions and scaling the external soft momenta, $k^{\mu}=\hbar\bar{k}^{\mu}$, we get
\begin{equation}
\begin{split}
    \mathcal{K}^{\mu}_{\text{soft}} &= Q_{1}^{3}Q_{2}^{3}e^{6}\int_{\omega}^{\omega +\delta\omega} d\Phi(k) k^{\mu}\int \prod_{i=1}^{2}\hat{d}^{4}q_{i}\ \hat{d}^{4}w_{i} \delta(2p_{i}\cdot q_{i})\delta(2p_{i}\cdot w_{i})\ e^{-iq_{1}\cdot b}\\
    & \hspace{5cm}\times\delta^{(4)}(w_{1}+w_{2})\delta^{(4)}(q_{1}+q_{2})\\
     &\hspace{-1cm}\bigg(\sum_{i=1}^{2}\bigg(-\frac{\varepsilon\cdot p_{i}}{p_{i}\cdot k}\bigg(\frac{w_{i}\cdot k}{p_{i}\cdot k}\bigg)+\frac{\varepsilon\cdot w_{i}}{p_{i}\cdot k}\bigg)\bigg)\bigg(\sum_{i=1}^{2}\bigg(-\frac{\varepsilon\cdot p_{i}}{p_{i}\cdot k}\bigg(\frac{(w-q)_{i}\cdot k}{p_{i}\cdot k}\bigg)+\frac{\varepsilon\cdot (w-q)_{i}}{p_{i}\cdot k}\bigg)\bigg)\ \frac{(4p_{1}\cdot p_{2})^{2}}{q_{1}^{2}(q_{1}-w_{1})^{2}}\ .
\end{split}
\end{equation}
Since the above argument is basically keeping track of the cancellation of factors of $\hbar$ arising from the amplitude and the measure, the conclusion that leading the contribution to the radiation flux comes from the $A^{\text{tree}}_{4+1}$ amplitude will hold for finite frequency radiation as well. The importance of the single photon emission amplitude has also been emphasized in the important work of \cite{Cristofoli:2021jas}, which we discuss now. 

\item  Since the out-state is computed perturbatively, we see that at a fixed order in the small deflection parameter, only a finite number of amplitudes with photons give the classical radiative flux. For instance, as we have seen at Leading order (LO - $e^{6}$), the $A_{5}^{\text{tree}}$ contributes. Similarly, at Next to Leading Order (NLO - $e^{8}$), we expect from power counting that the $A_{5}$ both at $1$-loop and tree-level contribute. Additionally, at each order, there can be a contribution from the product of tree amplitudes with photons, and these go as
\begin{equation}
    \begin{split}
        &e^{8}\rightarrow A_{4+2}^{*\text{tree}}A_{4+2}^{\text{tree}}\\
        &e^{10}\rightarrow A_{4+3}^{*\text{tree}}A_{4+3}^{\text{tree}}\\
        &e^{12}\rightarrow A_{4+4}^{*\text{tree}}A_{4+4}^{\text{tree}}\ \ \text{and so on}\ .
    \end{split}
\end{equation}
It might then seem that higher-order computations for the radiative flux would be a very hard task. However, in \cite{Britto:2021pud} it was shown that in the classical limit, all the above amplitudes with photons (or gravitons) are not independent and there exist relations between them. These relations were derived by assuming coherence of the outgoing radiation. Since the quantum state of the radiation can be described by a coherent state, the distribution of photons (or gravitons) will obey a Poissonian distribution. This means that the variance of the photon number operator should be equal to its mean in the classical limit. This is what leads to the set of relations between ``classical'' amplitudes with photons at each order in perturbation theory\footnote{It was assumed in their analysis that the super-classical terms coming from the amplitudes cancel. So by ``classical'' amplitude we mean that part of the amplitude which gives a non-trivial classical result.}. Similar relations can also be found by demanding the variance of operators be negligible in the classical limit, as has been shown in \cite{Cristofoli:2021jas}. We will now show that, as a consequence of these relations, the product of tree-level amplitudes with $X>1$ photons doesn't contribute to the classical result. For this, we shall again study the soft radiation flux. Let us start with the soft radiation flux, as defined in eq.\eqref{soft_rad_flux1} at leading order in soft frequency coming from the tree-level amplitudes. We get
\begin{equation}\label{tree_soft}
    \begin{split}
       \mathcal{K}^{\mu}_{\text{soft}} = &\sum_{X=1}^{\infty}\frac{1}{X!}\int_{\omega}^{\omega +\delta\omega}\prod_{i=1}^{X}d\Phi(k_{i})(\sum_{j=1}^{X}k_{j}^{\mu})\\
                             &\bigg| \int\prod_{i=1}^{2}\hat{d}^{4}q_{i}
                             \delta(2p_{i}\cdot q_{i}+q_{i}^{2})\ e^{-iq_{1}\cdot b/\hbar}\ \delta^{(4)}(q_{1}+q_{2})\\
                             &\hspace{3cm}\times\ (S^{(0)})^{X}\ A^{\text{tree}}_{4}(p_{1},p_{2}\rightarrow p_{1}+q_{1},p_{2}+q_{2})\bigg|^{2}\ .
    \end{split}
\end{equation}
We now take the classical limit and obtain
\begin{equation}
    \begin{split}
       \mathcal{K}^{\mu}_{\text{soft}} = &\sum_{X=1}^{\infty}\frac{\hbar^{2X+1}}{X!}\int_{\omega}^{\omega +\delta\omega}\prod_{i=1}^{X}d\Phi(k_{i})(\sum_{j=1}^{X}k_{j}^{\mu})\\
                             &\bigg| \hbar^{2}\int\prod_{i=1}^{2}\hat{d}^{4}q_{i}
                             \delta(2p_{i}\cdot q_{i})\ e^{-iq_{1}\cdot b}\ \delta^{(4)}(q_{1}+q_{2})\times\ \frac{1}{\hbar^{X/2}}(\bar{S}^{(0)})^{X}\ \frac{Q_{1}Q_{2}}{\hbar^{3}}\frac{(4p_{1}\cdot p_{2})}{q_{1}^{2}}\bigg|^{2}\\
                            &= \sum_{X=1}^{\infty}\frac{\hbar^{X-1}}{X!}\int_{\omega}^{\omega +\delta\omega}\prod_{i=1}^{N}d\Phi(k_{i})(\sum_{j=1}^{X}k_{j}^{\mu})\\
                             &\bigg| \int\prod_{i=1}^{2}\hat{d}^{4}q_{i}
                             \delta(2p_{i}\cdot q_{i})\ e^{-iq_{1}\cdot b}\ \delta^{(4)}(q_{1}+q_{2})\times\ (\bar{S}^{(0)})^{X}\ Q_{1}Q_{2}\frac{(4p_{1}\cdot p_{2})}{q_{1}^{2}}\bigg|^{2} .
    \end{split}
\end{equation}
Hence, we see that for $X>1$ all the contributions from the product of tree-level amplitudes vanish in the classical limit.\\
Returning to the analysis of \cite{Cristofoli:2021jas}, another important observation was that the relations derived there provided evidence that all information about the classical radiation flux is encoded entirely in the ``classical'' $A_{5}$ amplitude, namely the amplitude describing the emission of a single photon.
 An example of such a relation is the contribution of the product of two photon or graviton amplitudes at $1$-loop, to the radiative flux. This product is related to the product of five-point tree level amplitudes, written schematically 
\begin{equation}
    \hbar(A_{6}^{* 1-\text{loop}}A_{6}^{1-\text{loop}})=\frac{1}{2}\hbar(A_{5}^{*\text{tree}}A_{5}^{\text{tree}})^{2}.
\end{equation}
This might not be surprising given that a product of $n$-point functions of the radiative EM field in the coherent radiation out-state factorizes into a product of $1$-point functions, i.e.
\begin{equation}
    \bra{\text{out}}\mathbb{F}^{\mu_{1}\nu_{1}}(x_{1})\ldots \mathbb{F}^{\mu_{n}\nu_{n}}(x_{n})\ket{\text{out}} = \bra{\text{out}}\mathbb{F}^{\mu_{1}\nu_{1}}(x_{1})\ket{\text{out}}\ldots \bra{\text{out}}\mathbb{F}^{\mu_{n}\nu_{n}}(x_{n})\ket{\text{out}}\ .
\end{equation}
But it would be interesting to prove that the classical radiative flux for large impact parameter scattering can be computed solely from the knowledge of the ``classical limit'' of 5-point amplitudes. It is important to contrast this with the classical limit of the gravitational S-matrix. For this case, the (analog of) above-mentioned relations were studied in \cite{Britto:2021pud}. In this work, the authors reached the same conclusions as in the electromagnetic case for the linear gravitational memory. However, the recent study \cite{Georgoudis:2025vkk}  has shown that the $6$-point amplitude with two graviton emission does not vanish in the classical limit and in fact contributes to the non-linear gravitational memory\footnote{We thank Andrea Cristofoli for bringing our attention to this important point.}.  As the authors argue, this computation reveals that the quantum state of soft radiation at $\mathcal{I}^+$ is not a coherent state as it is sourced by the hard gravitons. It is the entanglement between soft and hard graviton modes that are responsible for non-linear memory from the perspective of the S-matrix. This should be contrasted with other radiative observables, such as the linear memory, e.g., which are sourced by classical objects. It will be interesting to compare this approach with our analysis in section \ref{gravity_sec}.
\item Finally, in the classical limit, quantum corrections are suppressed. This is simply because the Compton wavelength, $l_{c}\ll \sqrt{-b^{2}}$ as $\hbar\rightarrow 0$. Additionally, since the particles are described by wave packets, there is another length scale in the problem, the spread of the wave packets, $l_{w}$. For describing classical particles, we have $\hbar/l_{w}\ll m$, which says that the uncertainty in the momentum of the classical particles should be less than the masses of the particles. This implies $l_{w}\gg l_{c}$\footnote{For simplicity, we assume that the width of the wave packets in position space and the Compton wavelengths are the same for both the particles.}. So, the full set of inequalities to describe large impact parameter scattering is
\begin{equation}
l_{c}\ll l_{w} \ll \sqrt{-b^{2}}\ .
\end{equation}
The second inequality is chosen to make sure that the two initial particles are well separated. The above inequality is known as the Goldilocks hierarchy. We can now go back to the discussion of the scaling of massless momenta below eq.\eqref{rad_LIS}. Note that in the $\hbar\rightarrow 0$ limit, we could have taken $|\vec{b}|\rightarrow 0$ keeping $q^{\mu}$ finite. However, we see that in this limit the Goldilocks zone inequality is violated, and depending on the rate at which $(|\vec{b}|,\hbar)\rightarrow 0$, we can access the quantum regime of scattering. This would be the case when $|\vec{b}|\rightarrow 0$ faster than $\hbar\rightarrow 0$, and in this scenario, quantum corrections are dominant. The other possibility is $\hbar\rightarrow 0$ faster than $|\vec{b}|\rightarrow 0$. In this scenario, it is in principle possible to describe classical scattering as we can ensure that $l_{w}\gg l_{c}$, but the second inequality can be violated as $l_{w}\sim |\vec{b}|$ is a possibility in this case. 
\end{enumerate} 

We finally close this section with a few remarks.
\begin{enumerate}
\item In \cite{Caron-Huot:2023vxl}, the authors showed that using the on-shell S-matrix, one can construct an infinite hierarchy of out-of-time observables. The hierarchy is set by the number of out-of-time boundary insertions in the sequence of asymptotic operators. The inclusive observables computed via the KMOC formalism involve precisely one out-of-time insertion in the far future, and this class of observables was labeled $\textrm{Exp}_{3}$ observables. As an example of such an observable whose classical limit is the radiative electromagnetic field at ${\cal I}^{+}$ is the following, 
\begin{align}
\prod_{i=1}^{n} b_{\textrm{out}}^{(i)}(p_{i})\, a_{\textrm{in} \mu}(k) \prod_{j=1}^{m} b_{\textrm{out}}^{(j) \dagger}(p_{j}),
\end{align}
where $b(p)$ are matter operator and $a_{\mu}(k)$ is a photon operator. The subscripts out and in indicate that these operators are defined either in the far future or the far past, respectively. 
\item Even though the formalism has been very successfully applied to describe small-angle scattering via inclusive observables computed using the S-matrix, its applicability to small impact parameter scattering remains in its infancy. One of the notable exceptions is the seminal recent work by Aoki, Cristofoli, and Huang \cite{Aoki:2024boe}, where the authors analyze black hole mergers using the KMOC formalism  and show that so long as the ``Goldilocks hierarchy'' is satisfied, as
\begin{align}
l_{w}\, <<\, l_{s}\, <<\, GM(\textrm{Final mass after merger}) \,,
\end{align}
 the knowledge of three-point amplitude (BH + BH \, $\rightarrow\,$ merger) is enough to compute gravitational radiation in the soft limit and to the leading order in PM expansion. The reason why perturbative analysis applies to the case of merger is due to the fact that in \cite{Aoki:2024boe}, the final state in fact corresponds to the 2 particle state whose separation is less than the Schwarzschild radius of the total mass. The authors replace this two-particle state with a state of a single black hole with the associated form factor obtained via matching conditions. Their approach is very similar in spirit to the idea of ``stitching waveforms'' in the case of black hole binary inspiral. In that case, the Post-Newtonian wave form, which is valid until coalescence time, is stitched to the wave form computed using NR simulations. The ideas developed in \cite{Aoki:2024boe} have the potential for a more detailed analysis of black hole mergers beyond the leading PM expansion. 
\item However, we note that the classical scattering processes where exchange momenta are unbounded from above are outside the scope of their work, and as we show in the next section, thanks to the universality of multi-soft factorization theorems in quantum theory, soft radiative observables like memory can be computed using on-shell methods by extremizing relevant cross sections. Our goal in this paper is to bring these two ideas together in the context of electromagnetic and gravitational memory in $D = 4$ dimensions. 

\end{enumerate}

\section{Classical Soft theorem via Saddle point analysis}
\label{sec:classical_extremal}
In \cite{Laddha:2018rle}, an alternative approach was introduced for computing classical electromagnetic (and gravitational) radiation directly from the S-matrix. The basic idea is as follows. Given a set of semi-classical in and out states\footnote{An asymptotic single-particle state is specified by its momentum, spin, and any additional
quantum numbers such that, in the classical limit, it approximates a classical
object with corresponding asymptotic momentum, spin, and multipole moments.}, one evaluates the probability of emitting massless quanta into a detector bin
characterized by the frequency interval $[\omega, \omega + \delta\omega]$ and a
celestial angle $\Delta\Omega$. \emph{We will refer to such a detector
specification as a bin.} The natural question then is: for which configuration is this probability maximized?

The radiated flux is subsequently obtained by extremizing the semi-inclusive cross section with respect to the number of emitted quanta. This extremization determines the corresponding radiative field up to an overall phase. The practicality of this method in extracting classical observables directly from scattering amplitudes was demonstrated explicitly in the computation of low-frequency radiation. The key point is that if the soft factorization theorem holds for the emission of $X$ photons, i.e.,
\begin{align}
A_{4+X} \sim f_X\, A_4\, ,
\end{align}
then the extremization condition becomes independent of the hard amplitude $A_4$. Consequently, one obtains a direct relationship between the number of emitted photons, the asymptotic scattering momenta, and the detector parameters. Let us now summarize the result of their analysis.
\\ 
We restrict ourselves to electromagnetic scattering\footnote{The primary results in \cite{Laddha:2018rle} were obtained for a class of gravitational scattering processes, since graviton soft factorization theorems are more robust than their electromagnetic counterparts.}. Consider the classical scattering of $n \to m$ charged particles with charges $Q_i$ for $i=1,\ldots,n$ and $Q'_j$ for $j=1,\ldots,m$, and with incoming and outgoing momenta $p_1,\ldots,p_n$ and $p'_1,\ldots,p'_m$, respectively. The radiative gauge field associated with the low-frequency radiation emitted in this process, in spacetime dimension $D=4$, is
\begin{align}\label{34}
    \varepsilon_\mu A^\mu (\omega, \hat{n}) &= \frac{1}{2\pi i}\frac{e^{i\omega R}}{R}\ S_{em}^{(0)} (\varepsilon, k) \,,
\end{align}
where $k^\mu = \omega(1,\hat{n})$, to leading order in the soft frequency, $\omega$. Here $S_{em}^{(0)}$ is the leading soft factor, which depends only on the \textit{classical momenta} of the scattered particles and is 
\begin{align}
S^{(0)}_{\textrm{em}}(\varepsilon,k) = \sum_{i=1}^{m}Q'_{i} \frac{\varepsilon(k)\cdot p'_{i}}{p'_{i}\cdot k} - \sum_{j=1}^{n} Q_{j}\frac{\varepsilon(k)\cdot p_{j}}{p_{j} \cdot k} \,.
\end{align}
This result is known as the \emph{classical soft theorem}, and since it depends only on the asymptotic charges and momenta of the scattered particles, it is fully universal. Its robustness has been established by a direct analysis of the classical equations of motion for generic scattering processes \cite{Laddha:2019yaj}.

We now derive the above result in $\hbar \neq 1$ units. As discussed in the previous section, in this convention, both the charges and the outgoing soft-photon momenta carry explicit factors of $\hbar$. We will see that in the $\hbar\rightarrow 0$ limit, the extremization with respect to the number of emitted photons emerges naturally.

Consider the amplitude for the emission of $X$ soft photons each lying within the frequency bin $[\omega,\, \omega + \delta\omega]$ and angular bin $\Delta\Omega$ during the scattering of finite-energy particles with charges $Q_i$ and momenta $p_i$ into outgoing particles with charges $Q'_i$ and momenta $p'_i$:
\begin{align}
    p_1 + p_2+\ldots + p_{n} \rightarrow p'_{1} +\ldots + p'_{m} + X \, \text{soft photons}\,.
\end{align}
We take all soft photons with momentum $k$ to lie within the same bin. The relevant amplitude is then obtained from the multiple soft-photon theorem, evaluated at leading order in the soft frequency:
\begin{align}
    A_{n+m+X} (p_1,\ldots p_{n} \rightarrow p'_{1},\ldots p'_{m}, X \, k) & = (S_{em}^{(0)} (\varepsilon, k))^{X} A_{n+m} (p_1,\ldots p_{n} \rightarrow p'_{1},\ldots p'_{m})\,.
\end{align}
and $A_{n+m} $ is the amplitude of the process without soft photons. Then the differential cross section of emission of $X$ soft photons in the same bin is given by
\begin{align}
    P &= \frac{R^X}{X!} |A_{n+m}|^2 \,,  ~~~~~~\text{where}
 \, ~~   R = \frac{\omega \delta\omega \, \Delta \Omega}{2(2\pi)^3} \left|S_{em}^{(0)} (\varepsilon, k)\right|^2 \,.
\end{align}
Since the emitted photons are soft, we may pass to wavenumber space by writing $k^{\mu} = \hbar\, \bar{k}^{\mu}$ and rescaling the charges as $Q_i \rightarrow Q_i / \sqrt{\hbar}$. Taking the $\hbar \to 0$ limit, the expression above becomes
\begin{align}
    P &= \frac{\bar{R}^X}{X!\ \hbar^{X}} |A_{n+m}|^2 \,,  ~~~~~~\text{where}
 \, ~~   \bar{R} = \frac{\bar{\omega} \delta\bar{\omega} \, \Delta \Omega}{2(2\pi)^3} \left|\bar{S}_{em}^{(0)} (\varepsilon, \bar{k})\right|^2 \,.
\end{align}
where $(\bar{B})$ denotes that we have used the classical scaling for the quantity $B$. 
We now ask what the probability of emission of soft photons in this bin is, 
given that the in-out classical matter states are known. This is a conditional 
probability and is given by
\begin{equation}
     \mathcal{P}(\Delta \Omega,\omega,\delta \omega) = \frac{\sum_{X=0}^{\infty} X\ P}{|A_{n+m}|^2} = \lim_{\hbar\rightarrow 0} \sum_{X=0}^{\infty} \frac{\bar{R}^X}{X!\ \hbar^{X}} \ .
\end{equation}
To do this computation, we shall take the $\hbar \rightarrow 0$ limit first, 
before performing the sum. In this limit, the subsequent terms become more 
divergent in $1/\hbar$, and hence one must resort to a saddle point approximation 
with respect to $X$. However, we also note that the sum in the previous expression is constrained due to energy-momentum conservation and provides an upper bound on the number of photons emitted in this bin. Since $\Delta E:= E_{\text{in}}-E_{\text{out}}$, then the number of soft photons in this particular bin has to be such that $X^{*}(\hbar)< \frac{\Delta E}{\hbar \omega}$ where RHS is the maximum possible value of number of soft photons in this bin. So, the sum becomes 
\begin{equation}
     \sum_{X=1}^{\frac{\Delta E}{\hbar\bar{\omega}}} \lim_{\hbar\rightarrow 0} \frac{\bar{R}^X}{X!\ \hbar^{X}}.
\end{equation}
We now note that the sum should be dominated by a large value $X^{*}(\hbar)$ because the number of photons will scale with the charges of the classical matter particles. To compute this, we use the central limit theorem and use Stirling's formula to extremise with respect to $N$, the summand 
\begin{equation}
    \ln\bigg(\frac{\bar{R}^X}{X!\ \hbar^{X}}\bigg).
\end{equation}
So, we get
\begin{equation}
     X^{*}(\hbar) = \bar{R}/\hbar \ ,
\end{equation}
using which the energy (dominant part) emitted in the bin is 
\begin{equation}\label{rad_from_saddle}
    E^{*} = X^{*}\hbar \bar{\omega} = \bar{R}\ . 
\end{equation}
One can compare this to a direct classical computation for the low-frequency 
radiation. We know that the energy radiated away in a classical process can be 
written in terms of the radiative gauge field via the electromagnetic stress 
tensor. Comparing the classical computation for the low-frequency radiation 
emitted in a classical process to the above result, the radiative gauge field 
can be read off \cite{Laddha:2018rle}, and we obtain the classical soft theorem 
in eq.\eqref{34}. 
We now make a few remarks about the above derivation and compare it to the 
approach outlined in the previous section.

\begin{enumerate}
    \item The above calculation shows that the hard amplitude, $A_{n+m}$ doesn't play a role. This is because we are working in the in-out formalism, which is why we were able to ask a conditional probability. As a result, we also note that the infrared divergences dropped out because they were in the hard amplitude. Additionally, generically classical scattering will also have hard radiation (photons whose $\omega\sim E$), but since photons don't interact, we can compute the soft radiation flux in these processes as well using the above result. 
    \item We note that in contrast to the KMOC approach, the classical saddle in this approach is not dominated by the single photon emission but instead scales with the charge of the massive classical particles. Note that this is not in contradiction with the results obtained using the KMOC formalism. For large impact parameter scattering, the perturbation parameter is
    \begin{equation}
    \frac{Q_{1}Q_{2}}{E|\vec{b}|}\ll 1
    \end{equation}
    where $Q_{i}$s are the charges of the matter particles and $E$ is the typical centre of mass energy. So, we see that even though the charges of the individual classical particles can be large, the impact parameter $|\vec{b}|$ is large enough such that the above inequality is satisfied. 
    \item The main results of \cite{Laddha:2018rle} concern the classical soft theorems
for low-frequency gravitational radiation in $D>4$. As mentioned earlier, this
is because for $D>4$ the gravitational soft theorem is universal up to the
subleading order in frequency, $\mathcal{O}(\omega^{0})$. However, in computing
the subleading radiation, a subtlety arises. This originates from the fact that
the soft limit for multiple soft gravitons is not unique. When several soft
quanta are emitted, there are two inequivalent ways to take the soft limit.
The first possibility is to impose a specific ordering among the soft
frequencies, i.e.\ to take the soft limits at different rates. For example,
    \begin{equation}
        \omega_{1}>\omega_{2}>\ldots>\omega_{X} \ .
    \end{equation}
    The other possibility is to take all soft frequencies to zero at the same rate. Since we are interested in radiation for which all soft quanta lie in the same
detector bin, we work with the latter choice, known as the simultaneous soft
limit. However, in this case, the soft theorem for multiple soft gravitons at
subleading order in frequency does not, in general, factorize unless the
scattering is restricted to specific regimes, namely, large impact-parameter
scattering or the probe--scatterer approximation. Importantly, it was also
shown that restricting to these regimes ensures that the radiation emitted is
purely low-frequency.\\
Hence, we conclude that the extremization method for obtaining classical
low-frequency gravitational radiation, in $D>4$, from the S-matrix is valid only for these
two classes of classical processes.

\end{enumerate}
The saddle-point method computes the classical soft radiation flux for any
specified in-out matter state and does not rely on perturbation theory. This
implies that one can obtain the low-frequency radiative gauge field, to leading
order in the soft frequency, for a generic classical scattering process. This
stands in contrast to the KMOC formalism, where obtaining the same information
requires a complete resummation of the out state, including both the scattered
matter and the emitted radiation. Such a resummation is technically
challenging, even in the large impact-parameter regime.

\section{Leading classical soft photon theorem using KMOC paradigm}\label{main_sec}

As reviewed in the previous section, there is a prescription to compute classical radiation in soft expansion by extremizing the (conditional) probability distribution over the number of photons emitted in a given frequency bin.  The key ingredient in this argument is the equality of simultaneous and consecutive multi-soft limits (in the cases where the two differ at the quantum level, such as in non-Abelian gauge theories and gravity), which is equivalent to the vanishing of the contact terms in the classical limit.  As was discussed in the previous section, this is done by extremizing the probability of detecting $X$ photons (or gravitons) in a given bin at ${\cal I}^{+}$, given the asymptotic states of the ``hard particles''.  This result is independent of the details of the hard scattering.  Thanks to soft factorization, this probability is independent of the hard scattering amplitude, and as a result, classical soft radiation is completely fixed by quantum soft theorems.  

On the other hand, the basic premise of KMOC formalism is that if the incoming state $\vert \textrm{in} \rangle_{b}$ satisfies the so-called ``Goldilocks inequalities'', then one has a setup where the classical limit can be taken order by order in perturbative expansion of the amplitudes.  Thus, for any hyperbolic scattering process that is in the large impact parameter regime, the KMOC formalism can be used to compute radiation in any frequency bin order by order in perturbation theory. 

The key difference between the two formalisms in computing soft radiation is that the former depends on already specified initial and final states of the classical matter particles, whereas in the latter, the fate of the outgoing particles as well as the radiative field is computed perturbatively. 

But we would now like to see if at least the soft radiation can be computed as an inclusive observable, irrespective of the details of hard scattering.  More specifically, if we fix a bin at ${\cal I}^{+}$ that detects only soft radiation within a frequency band $[\omega,\, \omega + \delta\omega]$, then the multi-soft factorization of outgoing photons in this band implies the existence of a well-defined classical limit of the soft flux operator, independent of the details of the hard scattering. In other words, this classical limit persists even when the condition $\ell_{\omega}^{i} \ll \sqrt{-b^{2}} \;\forall\, i$ is violated. In this section, we will prove that this is indeed the case.

The inclusive observable we would thus like to compute is the following. Let, 
\begin{align}
\hat{K}^{\mu}_{\textrm{soft}} = \int_{{\cal B}} d\Phi(k)\, k^{\mu}\, \sum_{h = \pm} a_{h}^{\dagger}(k) a_{h}(k) \end{align}
where ${\cal B}$ is the integral over the bin. ${\cal B}$ is specified by the frequency in $[\omega, \omega + \delta \omega]$ with $\omega << m_{i}$ and is localized around a celestial direction $\hat{n}$ with an angular spread $\Delta  \Omega$.  

The associated in-in observable whose classical limit is the radiative flux in the KMOC formalism is then defined as 
\begin{align}
S^{\mu} :=\, \lim_{\hbar\, \rightarrow\, 0}\, \langle\psi \vert T^{\dagger} \hat{K}^{\mu}_{\textrm{soft}}\, T\, \vert\, \psi\rangle \,.
\end{align}
We remind the reader that we would like to compute this limit when the amplitude containing photons in the out-state in the prescribed bin factorizes according to leading soft factorization and with no assumptions on the perturbative expansion of the hard amplitude. 

\emph{We now consider an incoming state which violates the second inequality of the Goldilocks hierarchy} \cite{Kosower:2018adc}. That is, we consider an incoming coherent state of two charged scalar particles that scatter via a hard scattering operator with no parameter controlling the upper bound of exchange momenta.  Although the details of the scattering will not be relevant (owing to the universality of soft factorisation), it may be simpler to keep a specific model of scattering in mind, where the charged scalars interact via a short-range interaction of a massive vector boson with $M >> m_{i}$.   We then evaluate $S^{\mu}$ and show that it is well defined and finite, and compare the result with the classical limit of the multi-soft photon theorem computed in section \ref{sec:classical_extremal}. 

We thus start with an incoming two-particle coherent state with zero spatial separation.\footnote{Equivalently, we can also start with the state $\vert \textrm{in} \rangle_{b}$ and in the KMOC setup consider the limit where $\frac{\vec{b}}{\hbar}\, =\, \vec{\overline{b}}$ such that the exchange momenta do not scale with $\hbar$. We will comment on this limit at the end of this section.} 
\begin{align} \label{eq:zero_state}
\vert \textrm{in} \rangle_{0} :=\, \int \prod_{i=1}^{2} d\Phi(p_{i})\, \phi_{i}(p_{i})\ \vert p_{1}, p_{2}\, \rangle \,,
\end{align}
where $\phi_{i}(p_{i})$ is the gaussian wave packets defined in eq.\eqref{in_state}. We note that as
\begin{align}
\vert \textrm{in} \rangle_{0} :=\, \lim_{\vert \vec{b} \vert\, \rightarrow\, 0}\, \vert \textrm{in} \rangle_{b} \,, 
\end{align}
the hard amplitude involving scattering of the incoming state into a generic final state does not admit a perturbative expansion in $\frac{1}{\vert \vec{b} \vert}$. Thus, the only quantities we can hope to compute are those arising from soft limits of the amplitudes since soft theorems are exact statements in quantum field theory \cite{Sen:2017nim,Laddha:2017ygw,Laddha:2018rle,Laddha:2018myi,Laddha:2018vbn,Sahoo:2018lxl,Saha:2019tub,Laddha:2019yaj,Sahoo:2020ryf,Sahoo:2021ctw,Krishna:2023fxg,Sen:2024bax,Chakrabarti:2017ltl,AtulBhatkar:2018kfi,Ghosh:2021bam}.

We make one final comment before evaluating $S^{\mu}(\omega, \hat{n})$. 
\begin{itemize}
    \item In $D = 4$ dimensions, the S-matrix suffers from infrared divergence. Any observable which is infrared safe should then only depend on semi-inclusive cross section in which one traces over photons carrying non-zero momentum in the final state\footnote{In the KMOC formalism, such an infrared cancellation arising from the fact that classical observables are determined by semi-inclusive cross sections, can already be seen in the impulse at next-to-leading order in the electromagnetic coupling.}. In-in observables generically depend on semi-inclusive cross sections and hence, as we show below,  our final result for soft radiation in the classical limit will be infrared finite as it is determined by a specific semi-inclusive cross section \cite{Weinberg:1965nx}.
\end{itemize}

We now derive the formula for $S^{\mu}(\omega,\hat{n})$ to leading order in the
soft expansion. Let $A(p_1, p_2 \to R \cup X \cup Y)$ be the unstripped amplitude
for the two scalar particles with momenta $p_1,p_2$ and charges $Q_1,Q_2$ to scatter
into $R$ matter particles, $X$ photons inside the bin $\mathcal{B}$, and $Y$ photons
outside $\mathcal{B}$. Because all photons in $X$ are soft ($\omega_i\ll m_i$),
the amplitude factorizes at leading order via the multi-soft photon theorem. Moreover,
since the bin photons are soft, the momentum-conserving delta function depends only
on the total hard momentum $K_Y$ carried by photons outside the bin and is independent
of the soft momenta $k_i\in\mathcal{B}$:
\begin{equation}
  A(p_1,p_2 \to R \, \cup \, X \, \cup Y)
  \;=\;
  \delta^{(4)}\!\left(p_1+p_2 - P_R - K_Y\right)
  \left(\prod_{i=1}^{|X|} S^{(0)}_i\right)
  A_{2+|R|+|Y|}(p_1,p_2 \to R \, \cup \, Y)\,, \label{eq:soft_fact}
\end{equation}
where $S^{(0)}_i\equiv S^{(0)}_h(1,2,R,k_i)$ is the Weinberg soft factor for the
$i$-th bin photon and $A_{2+|R|+|Y|}$ is the amplitude for the same process without
the bin photons.
 
\smallskip
 
We evaluate $S^{\mu}$ by inserting a complete set of states in
$\mathcal{H}_{\mathrm{matter}}\otimes\mathcal{H}_{\mathrm{photon}}$. Splitting the
photon Hilbert space into photons inside the bin (set $X$) and outside (set $Y$),
the sum over intermediate states is an integral over multi-particle phase space
followed by a sum over the cardinalities $|X|$ and $|Y|$. Since
$\hat{K}^{\mu}_{\mathrm{soft}}$ acts only on photons in $\mathcal{B}$, it
annihilates any intermediate state with no photons in the bin, so the sum starts at
$|X|=1$:
\begin{align} 
  S^{\mu}
  &\;=\;
  \lim_{\hbar\to 0}
  \sumint_{\substack{|X|\geq 1\\|X|\in\mathcal{B}}}^{|X|_{\text{max}}(\hbar)}
  \;\sumint_{|Y|\geq 0}\;\sumint_R
  \;\;{}_0\langle\mathrm{in}|\,T^{\dagger}\,|R,X,Y\rangle
  \;\langle R,X,Y|\,\hat{K}^{\mu}_{\mathrm{soft}}\,T\,|\mathrm{in}\rangle_0\,.
\end{align}
The upper bound on the sum over $X$ is due to the following: the $|X|$ photons in the bin carry total energy
$E_X\ll\Delta E:=E_{\mathrm{in}}-(E_R+E_Y)$, so $|X|<\Delta E/\hbar\omega$; we
indicate this schematically as $|X|_{\max}(\hbar)$.
 
\smallskip
 
We now substitute the soft factorization eq.~\eqref{eq:soft_fact} to obtain
\begin{equation}
\begin{split}
    S^{\mu} = \lim_{\hbar\to 0}
  \sumint_{\substack{|X|\geq 1\\|X|\in\mathcal{B}}}^{|X|_{\text{max}}(\hbar)}&\sumint_{|Y|\geq 0}\;\sumint_R\ \langle \tilde{1},\tilde{2}|\,T^{\dagger}\,|R,Y\rangle
  \;\langle R,Y|\,T\,|1,2 \rangle\\
  & \big(\sum_{m=1}^{|X|} k_{m}^{\mu}\big) \big(\prod_{i=1}^{|X|}\sum_{h} S^{(0)}(1,2,R,k_{i}))\big(\prod_{j=1}^{|X|}\sum_{h'} S^{(0)}(1,2,R,k_{j})) 
\end{split}    
\end{equation}
We now note that since all bin photons share a narrow bin centred on $k_0$, we may approximate
$k_m^{\mu}\approx k_0^{\mu}$ and replace the phase-space integral over the $|X|$
photons by the bin volume $$|\mathcal{B}|=(|\vec{k}_0|^2\,\delta\omega)\Delta\Omega$$
raised to the power $|X|$. The product of $|X|$ soft factors together with their
complex conjugates, then takes the form
$\bigl(|S^{(0)}(1,2,R,k_0)|^2\bigr)^{|X|}$, where we use the condensed notation
\begin{equation}
  |S^{(0)}(\{p_i\},\{\tilde{p}_i\},\{r_j\}_{j\in R},k_0)|^2
  \;\equiv\;
  |S^{(0)}_{1,2,R,k_0}|^2
  \;:=\;
  \sum_{h}
  S^{(0)}_h(1,2,R,k_0)\,
  \sum_{h'} S^{(0)*}_{h'}(\tilde{1},\tilde{2},R,k_0)\,,
  \label{eq:notation}
\end{equation}
for the sum of squared soft factors. Here $S^{(0)}_h(1,2,R,k)$ is the Weinberg
soft factor
\begin{equation}
  S^{(0)}_h(1,2,R,k)
  \;=\;
  \sum_{j\in R} Q_j\,\frac{r_j\cdot\epsilon_h}{r_j\cdot k}
  \;-\;\sum_{i=1}^{2} Q_i\,\frac{p_i\cdot\epsilon_h}{p_i\cdot k}\,.
  \label{eq:weinberg}
\end{equation}
Assembling these ingredients, we obtain
\begin{align}
  S^{\mu}
  \;\approx\;
  \lim_{\hbar\to 0}
  \sum_R\;\sum_{|Y|\geq 0}
  \;\sigma(1 \, {+} \, 2 \, \to R\cup Y)\;
  \sum_{|X|=1}^{|X|_{\max}(\hbar)}
  \frac{|X|\,k_0^{\mu}}{|X|!}
  \Bigl(|\mathcal{B}|\,|S^{(0)}_{1,2,R,k_0}|^2\Bigr)^{|X|},
  \label{eq:before_sigma}
\end{align}
where $\sigma(1 \, {+} \, 2 \, \to R \, \cup \, Y)
$ is the product of the hard amplitude and its conjugate, 
smeared over the incoming wave packets as in eq.~\eqref{eq:zero_state}.
 
\smallskip
 
The final step is to trace over $Y$ by summing over all photons outside the bin.
This defines the semi-inclusive cross section smeared over incoming wave packets,
\begin{align}
  \sigma_{\mathrm{semi-inclusive}}(1 \, {+} \, 2 \, \to \, R)
  \;:=&\;
  \sum_Y\int
  \prod_{i=1}^{2}d\Phi_i(p_i)\,d\Phi_i(\tilde{p}_i)\,
  \phi_i(p_i)\,\phi^*_i(\tilde{p}_i)\,
  \delta^{(4)}\!\left(\sum_i p_i - \sum_i\tilde{p}_i\right)
  \notag\\
  &\times
  \delta^{(4)}\!\left(\sum_i p_i - K_Y - P_R\right)
  A(p_1{+}p_2\to R\cup Y)\,A^*({\tilde{p}_1 \, {+} \, \tilde{p}_2 \, \to R \, \cup \, Y})\,,  \label{eq:sigma_semi}
\end{align}
where $K_Y$ and $P_R$ are the total momenta carried by $Y$ photons and $R$ matter
particles respectively. Since the wave packets $\phi_i(p_i)$ are peaked around
the classical initial momenta $m_i u^{\mu}_i$, in the classical limit this reduces to
\begin{equation}
  \sigma_{\mathrm{semi-inclusive}}(1 \, {+} \, 2 \, \to \, R)
  \;=\;
  \sum_Y\delta^{(4)}\!\left(\sum_i p_i-K_Y-P_R\right)
  |A(p_1{+}p_2\to R\cup Y)|^2
  \;+\;\mathcal{O}(\hbar)\,. \label{eq:sigma_classical}
\end{equation}
We claim that $\sigma_{\mathrm{semi-inclusive}}\equiv\sum_Y\sigma_Y$ is infrared finite for
every $R$. This follows from considering the two possible cases. When
$E_{\mathrm{in}}-E_R>0$, summing over the photons in $Y$ that carry away the
missing energy necessarily includes a sum over all infrared photons of vanishing
energy; as a result, $Y$ is never empty, and $\sigma_Y$ is infrared finite. When
$E_{\mathrm{in}}=E_R$, the corresponding term in eq.~\eqref{eq:before_sigma} vanishes
because the action of $\hat{K}^{\mu}_{\mathrm{soft}}$ on the state $|R\rangle$
is trivial. Substituting eq.~\eqref{eq:sigma_classical}
into eq.~\eqref{eq:before_sigma} yields
\begin{align}
  S^{\mu}
  \;\approx\;
  \lim_{\hbar\to 0}
  \sum_R\;
  \sigma_{\mathrm{semi-inclusive}}(1 \, {+} \, 2 \, \to \, R)\;
  \sum_{|X|=1}^{|X|_{\max}(\hbar)}
  \frac{|X| k_0^{\mu}}{|X|!}
  \Bigl(|\mathcal{B}|\,|S^{(0)}_{1,2,R,k_0}|^2\Bigr)^{|X|}. \label{eq:kmusoftbin}
\end{align}

The departure from the KMOC formula arises from the fact that in the classical limit, we do not scale the exchange and loop momenta inside ${\cal A}_{2 + \vert R \vert + \vert Y \vert}$ with $\hbar$. Unlike the standard KMOC framework, which is naturally adapted to large impact-parameter scattering and hence small momentum transfer ($q \rightarrow 0$ or equivalently $q = \hbar \bar q$ with $\hbar \rightarrow 0$ and $\bar{q}$ held fixed), the hard-scattering regime considered here involves large momentum exchange. As a result, there doesn't exist a perturbative parameter and therefore admits neither a small-$q$ nor a small-$l$ scaling. 

\subsection{Classical limit of $\hat{K}^{\mu}_{\textrm{soft}}$} \label{sec:extremization}
We are finally in the position to compute the soft radiation flux $S^{\mu}$. For the benefit of the reader, we rewrite eq.(\ref{eq:kmusoftbin}) once again, 
\begin{align}
S^{\mu}\, \approx\,  \lim_{\hbar\, \rightarrow\, 0}\,   \sumint_{R}\, \sigma_{\textrm{semi-inclusive}}( 1 + 2\, \rightarrow\, R) \sum_{\vert X \vert = 1}^{\vert X \vert_{\textrm{max}}(\hbar)}\, \frac{\vert X \vert k_{0}^{\mu}}{\vert X \vert !}\, \big(\, \vert {\cal B}\, \vert \vert S^{(0)}_{1,2,R, k_{0}} \vert^{2} \big)^{\vert X \vert}.\nonumber\\
\end{align}
Let us isolate all the $\hbar$ scalings from the soft factor. Using the scaling described below eq.\eqref{rad_LIS} in section \ref{sec:kmoc}, it can be immediately seen that 
\begin{align}
 \vert {\cal B}\, \vert \vert S^{(0)}_{1,2,R, k_{0}} \vert^{2}\, =\, \frac{1}{\hbar}\, \overline{\vert {\cal B} \vert}\, \vert \overline{S}^{(0)}_{1,2, R, k_{0}} \vert^{2} \,. 
\end{align}
Hence, 
\begin{align}
S^{\mu}\, \approx\,  \lim_{\hbar\, \rightarrow\, 0}\,   \sumint_{R}\, \sigma_{\textrm{semi-inclusive}}( 1 + 2\, \rightarrow\, R)\, \sum_{\vert X \vert = 1}^{\vert X \vert_{\textrm{max}}(\hbar)}\, \hbar \vert X \vert \overline{k}_{0}^{\mu}\, \big(\, \frac{1}{\vert X \vert !}\, (\, \frac{\overline{\vert {\cal B} \vert} \vert \overline{S}^{(0)}_{1,2,R,\overline{k}_{0}}\, \vert^{2}}{\hbar}\, )^{\vert X \vert}\, \big) \,.
\end{align}
In the $\hbar\, \rightarrow\, 0$ limit, if there exists a saddle $\vert X \vert_{\star}(\hbar) >> 1$ which extremizes the summand, then the sum would be peaked around this value by the central limit theorem.  This expectation is rather reasonable since for a hard scattering with vanishing impact parameter, the number of photons emitted in the soft bin must scale with (classical) charges of the scattering particles.  We can find the extrema by extremizing log of the summand, $\ln\big(\, \frac{1}{\vert X \vert !}\, (\, \frac{\overline{\vert {\cal B} \vert} \vert \overline{S}^{(0)}_{1,2,R,\overline{k}_{0}}\, \vert^{2}}{\hbar}\, )^{\vert X \vert}\, \big)$.

The resulting saddle point equation can be solved using Stirling's formula and dropping subleading terms in $\frac{1}{\vert X \vert_{\star}(\hbar)}$. 
\begin{align}
\ln  \frac{\overline{\vert {\cal B} \vert} \vert \overline{S}^{(0)}_{1,2,R,\overline{k}_{0}}\, \vert^{2}}{\hbar}\, =\, \ln \vert X \vert_{\star}(\hbar)\nonumber\\
\implies\, \vert X_{\star} \vert(\hbar)\, =\,  \frac{\overline{\vert {\cal B} \vert} \vert \overline{S}^{(0)}_{1,2,R,\overline{k}_{0}}\, \vert^{2}}{\hbar} \,.
\end{align}
It can be immediately verified that the contribution of this saddle to $S^{\mu}$ is then given by, 
\begin{align}\label{softusingininkmoc}
S^{\mu}\, =\,  \sumint_{R}\, \big(\, (\lim_{\hbar\, \rightarrow\, 0} \sigma_{\textrm{semi-inclusive}}( 1 + 2\, \rightarrow\, R)\, ) \overline{\vert {\cal B}\vert}\, \vert \overline{S}^{(0)}_{1,2,R, k_{0}} \vert^{2}\, \overline{k}_{0}^{\mu} \, \big)
\end{align}
We thus see that for each set of outgoing momenta in $R, Y$ $S^{\mu}$ is proportional to the result obtained in eq.\eqref{rad_from_saddle} in section \ref{sec:classical_extremal}. The proportionality constant is the classical limit of the semi-inclusive cross section. 

Modulo the ``weight factor'' which is proportional to the semi-inclusive cross section in the classical limit, the above result appears to be closely related with the classical soft theorem, which states that given a fixed set of incoming and outgoing classical states (parametrized by electric charge and asymptotic momentum), the soft flux in the bin ${\cal B}$ equals
\begin{align}
S^{\mu}_{\textrm{classical-soft}}\, =\, \overline{\vert {\cal B}\vert}\, \vert \overline{S}^{(0)}_{\textrm{in}, \textrm{out}, k_{0}} \vert^{2}\, \overline{k}_{0}^{\mu} \,.
\end{align}
We can now go further and give a sharp equivalence between the two results. This equivalence is based on an argument that is partly heuristic, as the amplitudes themselves are infrared divergent. However, modulo such a caveat, this argument simply relies on our understanding of the classical limit of an S-matrix in Minkowski space-time \cite{Fabbrichesi:1993kz,Kim:2023qbl,Jain:2023fxc,Upadhyay:2025ged}. 

For simplicity, we consider a setup where $Y$ only contains essentially zero-frequency photons with $E_{Y} << \omega$ so that tracing over them simply cancels the virtual infrared divergence in $\sigma(1+2\, \rightarrow\, R)$.\footnote{We believe our arguments can be generalized to the case where the outgoing states also contain hard photons. However, as the hard photons themselves do not contribute to the Weinberg soft factor, we leave out this more general case for more detailed analysis in the future.}

Then, giving a fixed set of initial and final momenta in the classical theory is equivalent to the following constraint.
\begin{align}
\lim_{\hbar\, \rightarrow\, 0}\, {\cal A}(1+2\, \rightarrow\, \{r_{a}^{\textrm{out}}\})\, =\, e^{i S_{\textrm{on-shell}}( p_{1}, p_{2} \vert \{r_{a}^{\textrm{out}}\})}
\end{align}
where $S_\textrm{on-shell}$ is the on-shell action evaluated over the unique solution which is fixed by specifying incoming and outgoing momenta. As a result, we see that,
 \begin{align}\label{sicsclassicalf}
 \lim_{\hbar\, \rightarrow\, 0}\, \sigma_{\textrm{semi-inclusive}}(1 + 2\, \rightarrow\, R = \{ r_{a} \})\, =\, 
 \prod_{a = 1}^{\vert R \vert} \delta^{4}(r_{a} - r_{a}^{\textrm{out}})
 \end{align}
 where $\{r^{\textrm{out}}_{a}\}$ is a fixed set of outgoing momenta for all the particles in the set $R$.\footnote{This assumption is based on the realisation that when amplitudes are computed using path integrals, in the classical limit they peak around the saddle which is fixed by a set of incoming and outgoing momenta \cite{Fabbrichesi:1993kz,Jain:2023fxc}.}. Hence, the inclusive cross section in the classical limit would reduce to the RHS of eq.(\ref{sicsclassicalf}).

 Thus for $\forall\, \{r_{a}^{\textrm{out}}\}$, we have 
 \begin{align}\label{finalresult}
 S^{\mu} = S^{\mu}_{\textrm{classical-soft}}\, (\{p_{\textrm{in}}\}\, \rightarrow\, (\, \{r_{\textrm{out}}\}\, ) )
 \end{align}
 The sum over $R$ in eq.(\ref{softusingininkmoc}) is the sum over all the outgoing coherent states with fixed total momentum. Hence we can write $S^{\mu}$ formally as
 \begin{align}
 S^{\mu} = \sumint_{R\, \vert\, P^{\mu}_{R} = P^{\mu}_{in}}\, S^{\mu}_{\textrm{classical-soft}}(\{p_{in}\}\, \rightarrow\, R)
 \end{align}
 In summary, the soft radiation flux using the KMOC paradigm equals the RHS of eq.(\ref{softusingininkmoc}), and this result is a sum of classical soft factors over the momentum space of outgoing particles. 
 
 Some comments are in order.
\begin{enumerate}
\item Eq.(\ref{finalresult}) is consistent with the classical soft photon theorem since for every choice of outgoing matter state, the soft radiation is proportional to $\vert S^{(0)} \vert^{2}\, |\mathcal{B}|$ with weight one. 

 \item If the exchange momentum vanishes in the classical limit, as in the KMOC setup, the number $\vert X \vert$ of photons that contribute to the radiative flux in any bin $\langle\langle \hat{K}^{\mu} \rangle\rangle$ is fixed at each order in PL expansion. At the leading order in $e$, the contribution to the flux is obtained by summing over the set $X$ of all photons with $\vert X \vert = 1$. This is in sharp contrast with the analysis of the previous section, where $\vert X \vert$ is extremized in the classical limit.
\end{enumerate}

 \section{Gravitational memory as an Inclusive observable} \label{gravity_sec}
 In this section, we argue that the gravitational memory can also be computed as an inclusive observable for generic (as opposed to large impact parameter) scattering. Perhaps a concrete model for such a process is that of a black hole merger. Although the black-hole merger can also be analyzed in the KMOC formalism, where the final state is a 2-particle state in the Schwarzschild radius of the black hole at $i^{+}$, in our context, we simply consider a $2\, \rightarrow\, 1$ process which cannot be analyzed in perturbation theory. 
 
 Our primary result from the previous section is eq.\eqref{softusingininkmoc}, which we now summarize as follows. The leading order electro-magnetic soft radiation in a given bin, $\mathcal{B}$, can be written as, 
 \begin{align}\label{gnll}
S^{\mu} =\,  \sum_{R}\ \lambda_{R}\, \bar{\vert \mathcal{B}\vert} \vert S^{(0)}_{1,2,R, k_{0}}\vert^{2}k_{0}^{\mu}, 
 \end{align}
 where the weight $\lambda_{R}$ is (the classical limit of) the semi-inclusive cross section. That is, given the set of all semi-inclusive cross sections where the outgoing states $R$ have a fixed energy, we can compute soft flux using the in-in formalism. We will now argue that structurally this result does not hold in gravity, and this is due to the existence of a non-linear memory effect. While the main derivation mimics the EM case (and hence we do not repeat the details of extremization here), we show why in the gravity case, the result cannot be written in the form of eq.(\ref{gnll}). 
 
 Consider soft factorization of a gravitational amplitude where the set of incoming momenta $p_{1}, p_{2}$ scatters into a set of outgoing momenta $R\, \cup\, Y\, \cup\, X$ where $X$ denotes the set of soft gravitons in a given bin and $Y$ corresponds to the gravitons that carry the missing energy $E_{\textrm{in}}\, -\, E_{X} - E_{R}$. The soft factorization for the unstripped amplitude, for $\vert X \vert$ soft gravitons, all of which belong to the same bin ${\cal B}$, can be written as follows 
 \begin{align}
 {\cal A}_{2+ \vert R \vert + \vert X \vert + \vert Y \vert}\, =\, 
 \delta^{4}(P_{\textrm{in}}\, -\, P_{R} - K_{Y})\, 
 \frac{1}{\vert X \vert !}\, (\, S^{(0)}(1,2,R,Y \vert k_{0})\, )^{\vert X \vert}\, {\cal A}_{2 + \vert R \vert + \vert Y \vert} + \dots 
 \end{align}
 where the $\dots$ denote terms which are sub-leading in $k_{0}$ and
 \begin{align}\label{53}
 S^{(0)}(1,2,R, Y \vert k_{0})\, =\, 
 \sqrt{G_{N}}\epsilon_{\mu\nu}\, \bigg(\, \sum_{i=1}^{\vert R \vert}\, \frac{r_{i}^{\mu} r_{i}^{\nu}}{r_{i} \cdot k_{0}}\, -\, \sum_{j=1}^{2}\, \frac{p_{j \mu} p_{j \nu}}{p_{j} \cdot k_{0}}\, +\, \sum_{I=1}^{\vert Y \vert}\, \frac{k_{I}^{\mu} k_{I}^{\nu}}{k_{I} \cdot k_{0}}\bigg) \,.
 \end{align}
Note that every graviton in the set $Y$ is outside the bin in the sense that for $k^{\mu}\, \in\, Y$, $k \cdot k_{0}\, \neq\, 0$.

 The computation of 
$\langle\!\langle\, \hat{K}^{\mu}_{\mathrm{soft}}\, \rangle\!\rangle$
in the classical limit, now proceeds in complete analogy with the electromagnetic memory computation, including the extremization procedure discussed in Sec.~\ref{sec:extremization}, except that one now obtains
\begin{align}
S^{\mu}_{\mathrm{grav}}
\, =\,
\left(
\lim_{\hbar \rightarrow 0}
\sumint_{R \cup Y}
\sigma(1+2 \rightarrow R \cup Y)\,
\bigl|S^{(0)}(1,2,R,Y\,|\,k_{0})\bigr|^{2}
\,\overline{|{\cal B}|}
\right)
k_{0}^{\mu}\,,
\end{align}
where \(Y\) denotes the set of gravitons outside the chosen bin \({\cal B}\).

Now consider the subset of gravitons in \(Y\) whose total energy is of order \(O(\omega)\). We denote this subset by \(Y_{\omega}\), and its complement by
\((Y-Y_{\omega})\), which contains gravitons with energies much larger than \(\omega\). Unlike the electromagnetic case, the set \((Y-Y_{\omega})\) also contributes to the leading soft radiation due to the presence of the nonlinear (Christodoulou) memory effect.

The set \(Y_{\omega}\) also contains an unbounded number of infrared gravitons, and tracing over them therefore defines the semi-inclusive cross section
\[
\sumint_{Y_{\omega}}
\sigma_{1 \, + \, 2 \, \rightarrow R \, \cup \, Y_{\omega} \, \cup \, (Y-Y_{\omega})}\, .
\]
We note, however, that gravitons in \(Y_{\omega}\) do not themselves contribute to the nonlinear memory. Indeed, for a graviton with momentum
\(k_{1}\in Y_{\omega}\),
\begin{align}
\frac{k_{1}^{\mu}k_{1}^{\nu}}{k_{1}\!\cdot\! k_{0}}
\sim O(\omega^{0})\,,
\end{align}
so that these contributions are subleading in the soft expansion. Consequently,
\begin{align}\label{softgrav}
S^{\mu}_{\mathrm{grav}}
\, =\,
\left(
\lim_{\hbar \rightarrow 0}
\sumint_{R \, \cup \, (Y-Y_{\omega})}
\sigma_{Y_{\omega}}
(1 \, + \, 2 \, \rightarrow R \, \cup \, (Y-Y_{\omega}))
\,
\bigl|
S^{(0)}(1,2,R,(Y-Y_{\omega})\,|\,k_{0})
\bigr|^{2}
\,\overline{|{\cal B}|}
\right)
k_{0}^{\mu}\,,
\end{align}
where \(\sigma_{Y_{\omega}}\) denotes the infrared-finite semi-inclusive cross section.

We emphasize that \(S^{\mu}_{\mathrm{grav}}\) is infrared safe, with the hard gravitons in the set \((Y-Y_{\omega})\) contributing to the nonlinear memory at leading order in the soft expansion.

This formula is qualitatively different from its electromagnetic counterpart. In contrast to the electromagnetic memory formula, the gravitational expression receives additional contributions from hard gravitons in \(Y\), which source the nonlinear memory effect. In the probe--scatterer limit, the contribution of these gravitons to the Weinberg soft factor is suppressed by \(m/M\). In general, however, the formula requires: 
\begin{enumerate}
\item knowledge of the inelastic amplitudes, and
\item a separation between gravitons entering the chosen bin and those outside it.
\end{enumerate}
The latter set gives rise to nonlinear memory contributions to the leading soft radiation.

However, determining this contribution requires detailed information about inelastic scattering into hard gravitons, either through the semi-inclusive cross section or directly through the inelastic amplitudes themselves, owing to their dependence on hard gravitational radiation. Even with this information in hand, it remains unclear how to explicitly extract the nonlinear memory effect. The essential difficulty is that, unlike in the electromagnetic case, the classical limit of the inelastic amplitudes entering
\(
\sigma_{Y_{\omega}}
(1\, + \, 2 \, \rightarrow R \, \cup \, (Y-Y_{\omega}))
\)
is not a pure phase. Consequently, it is not immediately evident how the classical leading soft graviton theorem should emerge from this expression.

 \section{Conclusions}
 The non-perturbative aspect of quantum soft theorems is due to the fact that the incoming and outgoing states are independently specified in specifying an amplitude. The all-loop exactness of Weinberg soft theorem manifests itself in the all-order exactness of electromagnetic (or gravitational) memory, which is emitted in any electromagnetic (gravitational) scattering. It is then expected that an inclusive observable where we only specify the incoming state and compute the expectation value of radiative flux in a ``soft bin'' by tracing over all the outgoing states can leverage the non-perturbative exactness of the quantum soft theorems. However, the KMOC approach to classicality via inclusive observables relies on the validity of the PL (or PM) expansion of the QED (gravity) amplitude. In this paper, we have shown that one can stretch the KMOC formalism beyond the Goldilocks zone and still use it to compute soft electromagnetic (gravitational) radiation to all orders in the perturbative expansion. 
 
 Our final result for QED in the classical limit depends on the incoming momenta as well as outgoing momenta weighted by the (classical limit of the) corresponding semi-inclusive cross section, where one traces over all the photons outside the pre-specified bin. This result is consistent with the expectation that the complete physical information of the QED S-matrix in four dimensions is equivalent to specifying all possible semi-inclusive cross sections in the theory.\\ 
 For gravity, the situation is more subtle due to the presence of the non-linear memory effect. While soft gravitational radiation can be computed as an inclusive observable, we have not been able to recover the classical leading soft graviton theorem by taking the classical limit of the semi-inclusive cross section. This obstruction arises from the contribution of hard gravitational radiation.

 KMOC formalism has been used to show that in large impact parameter scattering, the classical limit of the quantum log soft theorem produces the universal log tail to subleading order in the PL (PM) expansion \cite{Georgoudis:2023eke,Paul:2026hyf}. In fact, in \cite{Paul:2026hyf}, the authors have shown that equating the classical limit of the quantum log soft theorem with the logarithmic tail in the radiative field at generic order in perturbative expansion implies some remarkable non-linear constraints on the S-matrix in the classical limit. However, neither the KMOC approach nor the method used in \cite{Laddha:2018rle} has shown the emergence of the classical log soft theorem from the quantum factorization to all orders in perturbation theory. It would be interesting to see if the ideas proposed in this work can be used to derive such a non-perturbative result from the S-matrix. 
 
\acknowledgments
AL would like to thank Miguel Campiglia, Sangmin Choi, Andrea Puhm, and Ashoke Sen for insightful discussions on the relationship between classical and quantum soft theorems. We would also like to thank Siddhartha Paul and Adarsh Vishwakarma for sharing their results prior to publication and for several clarifications about the status of log soft theorems in the KMOC formalism. SA is supported by the São Paulo Research Foundation (FAPESP) under Grant No. 2025/01291-6.

\bibliographystyle{JHEP}

\bibliography{collision}

\end{document}